\documentclass[aps,twocolumn,showpacs,superscriptaddress]{revtex4}
\usepackage{epsfig}
\usepackage{dcolumn}
\begin{document}


\title{Anisotropic magnetoresistive and magnetic properties of
La$_{0.5}$Sr$_{0.5}$CoO$_{3-\delta}$ film}

\author{B. I. Belevtsev}
\email[]{belevtsev@ilt.kharkov.ua}
\affiliation{B. Verkin Institute for Low Temperature Physics and Engineering,
National Academy of Sciences, Kharkov 61103, Ukraine}

\author{V. B. Krasovitsky}
\affiliation{B. Verkin Institute for Low Temperature Physics and Engineering,
National Academy of Sciences, Kharkov 61103, Ukraine}

\author{A. S. Panfilov}
\affiliation{B. Verkin Institute for Low Temperature Physics and Engineering,
National Academy of Sciences, Kharkov 61103, Ukraine}

\author{I. N. Chukanova}
\affiliation{Institute for Single Crystals, National Academy of Sciences,
Kharkov 61001, Ukraine}



\begin{abstract}
The magnetic and transport properties of La$_{0.5}$Sr$_{0.5}$CoO$_{3-\delta}$
film grown on a LaAlO$_3$ substrate by pulsed-laser deposition are
studied. The properties are found to be influenced by the magnetic
anisotropy and inhomogeneity. Magnetoresistance anisotropy is determined by
the shape anisotropy of the magnetization and the strain-induced
magnetic anisotropy due to the film-substrate lattice interaction.
Indications of the temperature-driven spin reorientation transition from
an out-of plane orderded state at low temperatures to an in-plane
ordered state at high temperatures as a result of competition between the
mentioned sources of magnetic anisotropy are found.
\end{abstract}


\maketitle

\section{Introduction}
Mixed-valence lanthanum cobaltites of the type  La$_{1-x}$Sr$_{x}$CoO$_{3}$
have attracted much attention in recent years due to their unique magnetic
and transport properties \cite{itoh,gooden}. Study of this system is also
important for understanding the nature of colossal magnetoresistance
in the related oxides, mixed-valence manganites \cite{coey,dagotto}.
For technical application, the epitaxial films of these compounds
are mainly implied to be used. In that case the shape anisotropy
(due to the demagnetization effect) and the
film-substrate lattice interaction can induce magnetization anisotropy
and, therefore, magnetoresistance (MR) anisotropy (bulk samples of these
compounds show no marked magnetic or MR anisotropy). This point
was  studied rather intensively in manganite films (see \cite{belev1} and
references therein). Studies of this type can hardly be found in
literature for cobaltites. In addition, the properties of mixed-valence
cobaltites  are influenced by their unavoidable magnetic inhomogeneity,
which is caused by different extrinsic and intrinsic reasons.
The extrinsic ones are determined by various technological factors in the
sample preparation. They can cause inhomogeneity in chemical composition
(for example in oxygen concentration) or in crystal structure
(polycrystalline or granular samples). The intrinsic sources of
inhomogeneity are believed to arise for thermodynamical reasons and can
lead to phase separation into two phases with different concentration of
the charge carriers and, therefore, to significant magnetic inhomogeneity
\cite{itoh,gooden,belev2}. In this article we
present a study of La$_{0.5}$Sr$_{0.5}$CoO$_{3-\delta}$ film which
demonstrates a combined influence of the magnetic anisotropy and
inhomogeneity on its transport, magnetoresistive and magnetic properties.
Indications of the temperature-driven spin reorientation transition from
an out-of plane orderded state at low temperatures to an in-plane
ordered state at high temperatures as a result
of competition between the mentioned anisotropy sources are found.

\section{Experimental}
The La$_{0.5}$Sr$_{0.5}$CoO$_{3-\delta}$ film (about 220 nm thick)
was grown by pulsed-laser deposition (PLD) on a (001) oriented LaAlO$_3$
substrate. The ceramic target used was prepared by a standard solid-state
reaction technique. A PLD system with an Nd-YAG laser operating at 1.06
$\mu$m was used to ablate the target. The pulse energy was about 0.39 J with
a repetition rate  of 12 Hz and pulse duration of 10 ns.
The film was deposited with a substrate temperature of 880$\pm 5^\circ$C
in oxygen atmosphere at a pressure of about 8~Pa. The film was cooled
down to room temperature after deposition at an oxygen pressure about 
$10^5$~Pa. The target and film were characterized by X-ray diffraction 
(XRD) study. 
\par
The film resistance, as a function of temperature and magnetic field
$H$ (up to 20 kOe), was measured using a standard four-point technique. 
The field was applied parallel or perpendicular to
the film plane. In both cases it was perpendicular to the transport 
current. The magnetization, $M$, was measured in a Faraday-type 
magnetometer. A rotating electromagnet makes it possible to measure
the magnetization with different directions of $H$ relative to the
plane of the film.

\section{Results and discussion}

\begin{figure}[htb]
\centerline{\epsfig{file=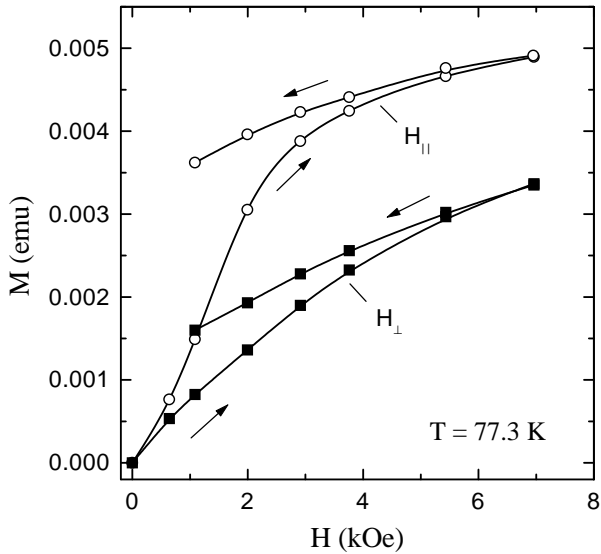,width=8.5cm}}
\caption{Magnetization curves of the film studied for the fields parallel
($H_{\parallel}$) and perpendicular ($H_{\perp}$) to the film plane.}
\label{fig1}
\end{figure}

\begin{figure}[htb]
\centerline{\epsfig{file=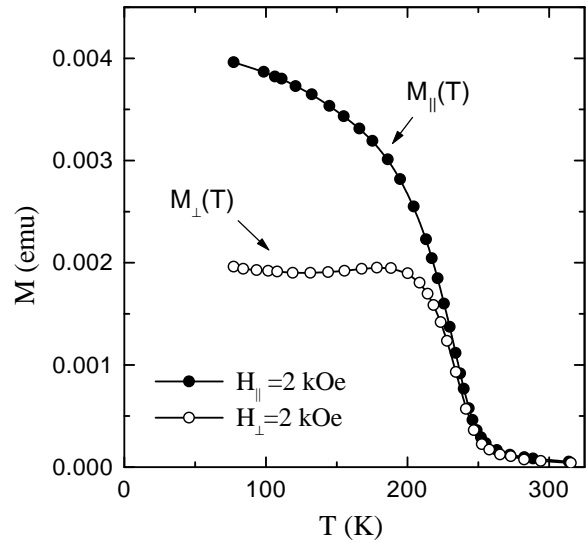,width=8.5cm}}
\caption{Temperature dependences of the magnetization of the film studied 
for the magnetic field ($H=2$~kOe) applied parallel [$M_{\parallel}(T)$] and 
perpendicular [$M_{\perp}(T)$] to the film plane. 
The thermomagnetic prehistory: the sample was cooled down to liquid nytrogen
temperature, $T\approx 77.3$~K, 
in a field close to zero, then field was increased up to 7 kOe and lowered down
to 2 kOe (see Fig. \ref{fig1}). After that the dependences have been 
recorded at that field with temperature increasing.}
\label{fig2}
\end{figure}
\par
We have found a strong anisotropy in magnetic and magnetoresistive 
properties of the film studied. The anisotropy manifests itself as 
dramatic differences in those properties for magnetic fields applied 
parallel and perpendicular to the film plane. Consider at first the 
anisotropy of magnetic properties. Magnetization curves for the fields 
parallel ($H_{\parallel}$) and perpendicular ($H_{\perp}$) to the film 
plane demonstrate a strong anisotropy (Fig. \ref{fig1}). At the maximum 
field applied (7 kOe), the magnetization seems to be rather close to 
saturation for the in-plane field orientation, but it is far from it 
for the out-of-plane one.  It is reasonable to suppose that this is 
determined mainly by the shape anisotropy.  
\par
Temperature dependences of the film magnetization for the field directions 
parallel [$M_{\parallel}$] and perpendicular [$M_{\perp}$] to the film 
plane are shown in Fig. \ref{fig2}. The Curie temperature, $T_{\mathrm{c}}$, 
is found to be about 250~K. The $M_{\parallel}(T)$ behaviour is quite common 
for ferromagnetic (FM) metals: it saturates with temperature decreasing. 
The behavior of $M_{\perp}(T)$ is quite different from that of 
$M_{\parallel}(T)$. At fairly high field used, 2~kOe, the $M_{\perp}(T)$ 
curve is found to be well below the $M_{\parallel}(T)$ curve. Besides, in 
low temperature range the $M_{\perp}(T)$  curve is non-monotonic 
(Fig. \ref{fig2}).
\par
\begin{figure}[htb]
\centerline{\epsfig{file=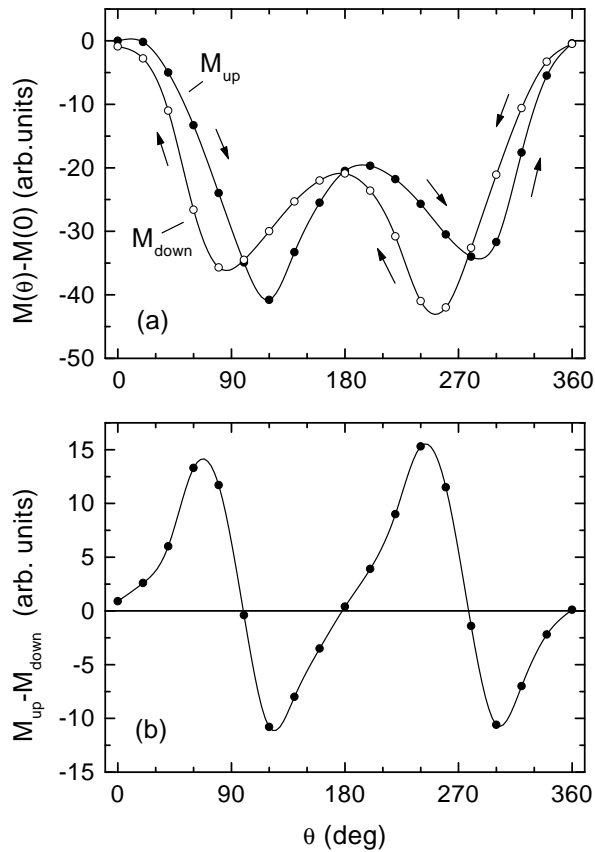,width=8.5cm}}
\caption{Panel $(a)$ presents dependences of the magnetization on the 
angle $\theta$ between the magnetic field and the film plane 
(at $H=2$~kOe and $T=77.3$~K). The thermomagnetic prehistory is described
in capture for Fig. \ref{fig2}. The curves $M_{\mathrm{up}}(\theta)$ and 
$M_{\mathrm{down}}(\theta)$ were recorded with step rotating of the field 
from 0$^{\circ}$ to 360$^{\circ}$ and back to 0$^{\circ}$, respectively.  
It can be seen a considerable hysteresis effect. The angular dependence of 
the difference between the $M_{\mathrm{up}}(\theta)$ and 
$M_{\mathrm{down}}(\theta)$ in panel $(b)$ enables to present this effect 
more clearly.}
\label{fig3}
\end{figure}

 
Figure \ref{fig2} presents actually the $M(T)$ behavior only for  
two values of the angle, $\theta$, between the field and the film plane: 
$\theta=0^{\circ}$ and $\theta=90^{\circ}$. It is helpful to consider 
the whole angle dependences of the magnetization which are presented
in Fig. \ref{fig3}(a). Here, the curves $M_{{\mathrm{up}}}(\theta)$ and 
$M_{{\mathrm{down}}}(\theta)$ were recorded with step rotating of the field 
from 0$^{\circ}$ to 360$^{\circ}$ and back to 0$^{\circ}$, respectively.  
It can be seen that the magnetization takes maximum values at 
$\theta \approx$~0$^{\circ}$, 180$^{\circ}$ and 360$^{\circ}$, that 
is for the in-plane field orientations. The magnetization magnitude
at $\theta \approx$~180$^{\circ}$ is less than these at  
$\theta \approx$~0$^{\circ}$ and 360$^{\circ}$. This is determined by the 
shape of a magnetization loop and thermomagnetic prehistory of the sample.
The minimum magnetization values are found, as expected, at 
$\theta \approx$~90$^{\circ}$ and 270$^{\circ}$, that is for the 
out-of-plane field orientations. 
\par
It is seen a considerable hysteresis effect in the $M(\theta)$ curves
[Fig. \ref{fig3}(a)]. To present the effect more clearly, an angular 
dependence of the difference between the $M_{\mathrm{up}}(\theta)$ and 
$M_{\mathrm{down}}(\theta)$ is shown in Fig. \ref{fig3}(b). The function
$d(\theta)=M_{\mathrm{up}}(\theta)-M_{\mathrm{down}}(\theta)$ can be taken 
as some measure of the angular hysteresis effect. It is seen that 
the $d(\theta)$ dependence is close to a periodic one with a period equal 
to $180^{\circ}$. It takes zero value at the angles multiple of $90^{\circ}$, 
corresponding to both the in-plane and out-of-plane directions of magnetic 
field [Fig. \ref{fig3}(b)]. The extreme values of $d(\theta)$ are situated
at some intermediate angles, which are, however, more close to the 
out-of-plane directions than to the in-plane ones. 
\par
As indicated above, the magnetization anisotropy in the film studied should 
be determined mainly by the shape anisotropy. Closer inspection shows, 
however, that  $M_{\perp}(T)$ behavior cannot be attributed solely to the 
shape-anisotropy effect: $M_{\perp}(T)$ and $M_{\parallel}(T)$ are 
practically equal in rather broad temperature range just below 
$T_{\mathrm{c}}$, then (going to lower temperature) the $M_{\perp}(T)$ curve 
goes rather abruptly well below the $M_{\parallel}(T)$ curve and becomes 
non-monotonic with a pronounced increase in $M_{\perp}(T)$ at low 
temperatures (Fig.~\ref{fig2}). These $M_{\perp}(T)$ features can be caused 
by the strain-induced magnetic anisotropy due to lattice mismatch between the 
film and the substrate. 
This guess is supported by our XRD study which has revealed that the film has 
an out-of-plane tensile strain. For materials with the positive
magnetostriction this must favor an out-of-plane easy magnetization. 
An additional corroborations of this suggestion have been found in  
the MR properties of the film, described below.

\par
\begin{figure}[htb]
\centerline{\epsfig{file=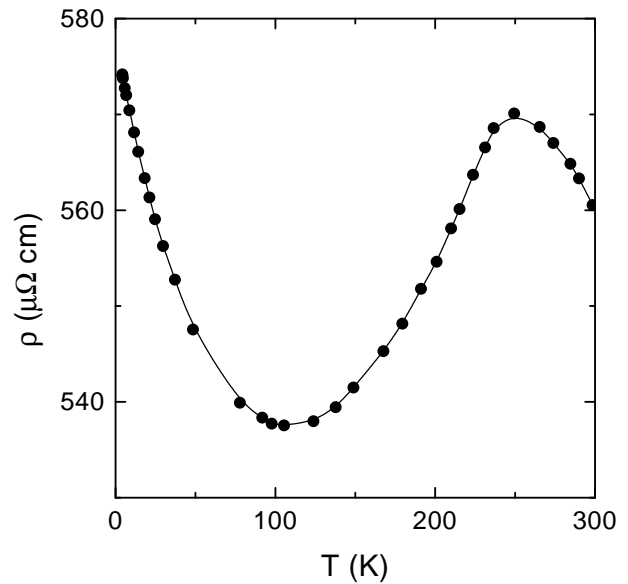,width=8.5cm}}
\caption{Temperature dependence of the film resistivity.}
\label{fig4}
\end{figure}

Now turn to transport properties of the film. The temperature dependence of 
the resistivity, $\rho (T)$, is found to be non-monotonic (Fig. \ref{fig4}) 
with a maximum at $T\approx 250$~K and a minimum at $T\approx 107$~K.  
La$_{0.5}$Sr$_{0.5}$CoO$_{3-\delta}$ samples with fairly perfect 
crystalline structure and $\delta$ close to zero are  known to be metallic 
(${\mathrm{d}}\rho/{\mathrm{d}}T > 0$) in the whole range below and above
$T_{\mathrm{c}}$ \cite{gooden}. The $\rho (T)$ behavior in Fig. \ref{fig4} 
reflects inhomogeneous structure of the film and some oxygen deficiency. Due 
to the last factor, the hole concentration is less then a nominal one 
(at $\delta=0$). This is responsible for a resistance peak at $T=250$~K 
which is common for low-doped La$_{1-x}$Sr$_{x}$CoO$_{3}$  with 
$0.2 \leq x \leq 0.3$ \cite{gooden}. The low temperature resistance minimum  
is typical for systems of FM regions (grains or clusters) with rather weak 
interconnections. For example, it has been frequently seen in polycrystalline 
manganites \cite{mahen,andres,auslender}. The inhomogeneous structure can 
be determined by technological factors of sample preparation (causing the 
polycrystalline structure with rather high tunneling barriers between the 
grains) or by the phase separation into the hole-rich and hole-poor phase 
\cite{itoh,gooden}. The conductivity of inhomogenenous systems of this type 
is determined by the intragrain conductivity and the tunneling of charge 
carriers through the boundaries between the grains. A competition between
these two contributions can lead to a resistance minimum 
\cite{andres,auslender}. For an extended discussion of the most obvious 
reasons for the appearance of the resistance minimum in polycrystalline
cobaltites see Ref. \cite{belev2}. 

\par
\begin{figure}[htb]
\centerline{\epsfig{file=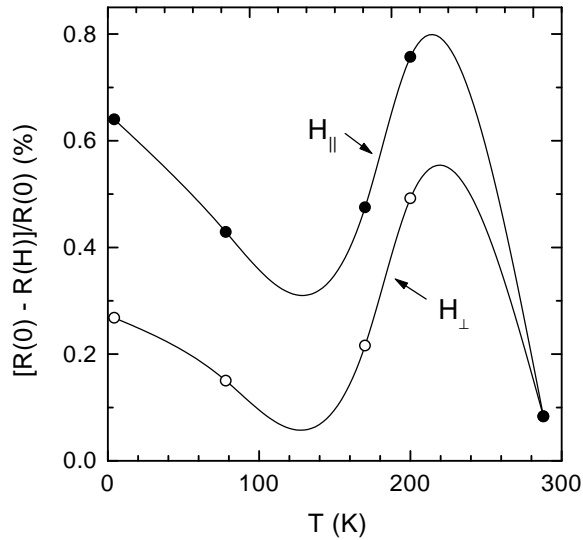,width=8.5cm}}
\caption{Temperature dependence of the magnetoresistance at $H=20$~kOe
for fields parallel ($H_{\parallel}$) and perpendicular ($H_{\perp}$) to the 
film plane. In both cases the fields were perpendicular to the transport
current. The solid lines present a B-spline fitting.}
\label{fig5}
\end{figure}
\begin{figure}[htb]
\centerline{\epsfig{file=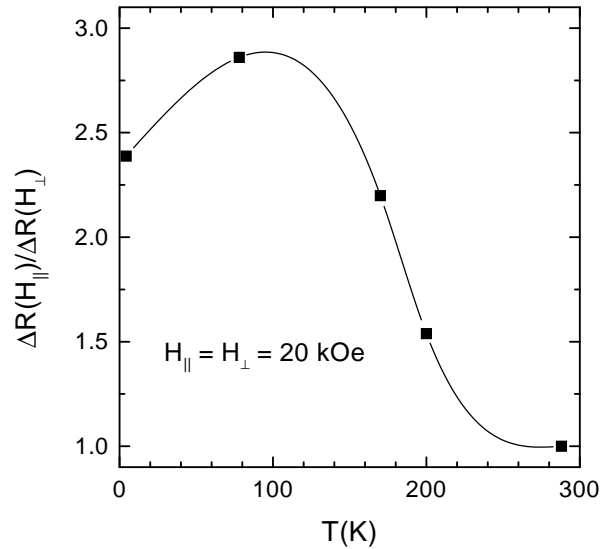,width=8.5cm}}
\caption{Temperature dependence of the ratio of magnetoresistances for
fields parallel ($H_{\parallel}$) and perpendicular ($H_{\perp}$) to 
the film plane. The fields were equal to 20 kOe.}
\label{fig6}
\end{figure}

The MR in the film studied is found to be anisotropic. The absolute values 
of negative MR in fields parallel to the film plane are considerably above 
those in perpendicular fields (Fig. \ref{fig5}). The temperature behavior of 
the ratio between the in-plane and out-of-plane MRs is shown in 
Fig. \ref{fig6}. It is seen from Figs. \ref{fig5} and \ref{fig6} that this 
MR anisotropy takes place only in FM state and disappears for
$T>T_{\mathrm{c}}$.
Since the conductivity of mixed-valence cobaltites increases with
enhancement of the magnetic (spin) order, this behavior just reflects the
point that the magnetization increases more easily in a magnetic field
parallel to the film plane, as it has been indeed found in this study
(Figs. \ref{fig1}, \ref{fig2} and \ref{fig3}).

\par
In polycrystalline samples (beside an intrinsic MR, which depends on
magnetic order inside the grains) a significant contribution to the MR
comes from grain boundaries, and this contribution increases with decreasing
temperature. Discussion of the possible mechanisms for this extrinsic type of
MR can be found in Refs. \cite{gupta,hwang,evetts,ziese}. The film studied
shows indeed a continuous increase in MR for decreasing temperature
(for the temperatures well below $T_{\mathrm{c}}$) (Fig. \ref{fig5}). This
behavior is expected for polycrystalline FM samples with poor enough
intergrain conductivity \cite{gupta,hwang}. In contrast, for cobaltite and
manganite samples with fairly good crystal perfection and even for
polycrystalline samples of these materials but with a good intergrain
connectivity, the MR goes nearly to zero with decreasing temperature
\cite{gupta,yamaguchi}. It should be mentioned that grain boundaries in 
FM oxides are regions of perturbation of structural and magnetic orders, 
and, therefore, induce a magnetic  inhomomogeneity as well. These boundaries 
(and, maybe, other sources of inhomogeneity, i.e., the phase separation) 
can cause the significant angular
hysteresis effect found in this study (Fig. \ref{fig3}) since they hinder
the motion of FM domains at a rotation of magnetic field. It is noteworthy,
however, that hysteresis effect is minimal at the angles corresponding to
both the in-plane and out-of-plane directions of magnetic field.
In summary, the behavior of resistivity, MR and magnetization of
the film corresponds to that of a system of weakly connected grains.

\begin{figure}[htb]
\centerline{\epsfig{file=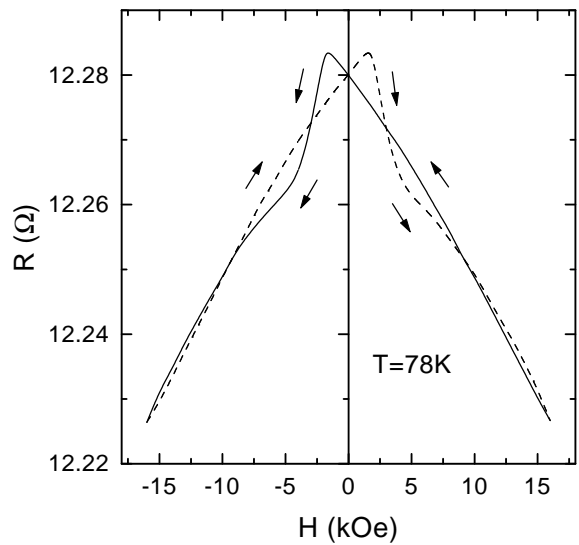,width=8.5cm}}
\caption{Magnetoresistive hysteresis at $T=78$~K for fields parallel to
the film plane and perpendicular to the transport current.}
\label{fig7}
\end{figure}

\begin{figure}[htb]
\centerline{\epsfig{file=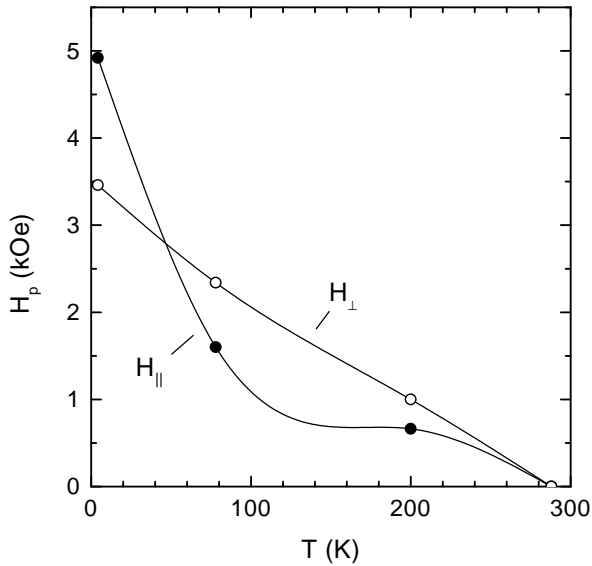,width=8.5cm}}
\caption{ Temperature dependence of characteristic field, $H_{\mathrm{p}}$,
at which resistance peaks in the magnetoresistive hysteresis curves 
(Fig. \ref{fig7}), for 
fields parallel ($H_{\parallel}$) and perpendicular ($H_{\perp}$) to 
the film plane. The field $H_{\mathrm{p}}$ corresponds to coercive force 
($H_{\mathrm{c}}$) in magnetization loops.}
\label{fig8}
\end{figure}

\par
The data presented in Fig. \ref{fig5} are pertaining to negative MR for
fairly high fields. In general, the MR curves are hysteretic and have
specific features in low-field range (Fig. \ref{fig7}). Symmetric
hysteresis curves, like that in Fig. \ref{fig7}, were obtained 
for the film studied after some number of repeated sweeps between the 
chosen maximum (positive and negative) field magnitudes. For the first
sweeps, the hysteresis curves were somewhat asymmetric. Actually, their 
behavior correlates with that of magnetization loops \cite{ziese}. In 
particular, the field $H=H_{\mathrm{p}}$, at which resistance peaks 
(Fig. \ref{fig7}), corresponds to value of the coercive force 
($H_{\mathrm{c}}$). The value of $H=H_{\mathrm{p}}$ decreases with 
increasing temperature and goes to zero with approaching $T_{\mathrm{c}}$. 
The magnitude of positive MR in the low-field range, 
$\Delta R(H_{\mathrm{p}}) = [R(H_{\mathrm{p}})-R(0)]/R(0)$, is some 
measure of the remanent magnetization. 
\par
We found that $H_{\mathrm{p}}$ and $\Delta R(H_{\mathrm{p}})$ depend on the 
field direction and reflect in this way the magnetization anisotropy.
The temperature dependences of $H_{\mathrm{p}}$ for the in-plane and
out-of-plane directions of magnetic field are shown in Fig. \ref{fig8}.
It is seen that at $T\simeq 4.2$~K the value of $H_p$ in the out-of-plane
field is less than that in the in-plane field, but at $T\approx 78$~K and
higher temperatures the opposite relation is true. For high enough
temperature ($T>T_{\mathrm{c}}$) the  $H_{\mathrm{p}}$ values go to zero for
both field directions. The $\Delta R(H_{p})$ values are found to be higher
for the out-of-plane field direction as compared with the in-plane one at
$T\simeq 4.2$~K. At $T\approx 78$~K and $T=200$~K, the opposite relation
holds true. All this implies that at low temperatures the out-of-plane
magnetization is favored, whereas for higher temperatures the in-plane
magnetization becomes dominant. The pronounced increase in $M_{\perp}(T)$ at
low temperatures (Fig. \ref{fig2}) and decrease in the ratio between the 
in-plane and out-of-plane MRs  below $T\approx 80$~K (Fig. \ref{fig6}) 
support additionally this suggestion. All these are indications of the
temperature-driven spin reorientation transition which can be determined by 
competition between the shape anisotropy and the strain-induced anisotropy.
This transition has been studied rather intensively (theoretically and
experimentally) for films of common FM metals \cite{hucht,hu}, but has
never been mentioned for cobaltite films. It should be noted, however, that
theoretical models, like \cite{hucht,hu}, are applicable only for ultrathin
magnetic films (up to 10 monolayers), whereas the film studied is much
thicker and rather disordered. Consequently, a spin-reorientation transition 
in the film studied can have a different nature than those proposed for 
ultrathin films.
\par
In conclusion, we have revealed and investigated the magnetic and
magnetoresistance anisotropy in La$_{0.5}$Sr$_{0.5}$CoO$_{3-\delta}$ film.
Among other things, we found indications of the temperature-driven spin 
reorientation transition in the film studied:
at low temperature, the magnetization vector is prone to be perpendicular
to the film plane, but by increasing the temperature the magnetization vector
goes entirely to the in-plane direction.

\end{document}